\documentclass[twocolumn,
    preprintnumbers, superscriptaddress, amsmath, amssymb, prd, reprint, floatfix, nofootinbib ]{revtex4-2}

\usepackage{graphicx}
\usepackage[dvipsnames]{xcolor}
\usepackage{amsfonts}
\usepackage{bm}
\usepackage{slashed}
\usepackage[normalem]{ulem}
\usepackage{url}
\usepackage{booktabs}
\usepackage{hyperref}
\hypersetup{
  colorlinks=true, linkcolor=blue, citecolor=blue, filecolor=black, urlcolor=blue, }

\setcitestyle{square,numbers,sort&compress}

\begin{document}

\preprint{ADP-26-15/T1312}

\title{Endothermic dark matter with a light dark photon and the
LUX--ZEPLIN high-energy nuclear-recoil candidate}

\author{Pengxuan Zhu}
\email{pengxuan.zhu@adelaide.edu.au}
\affiliation{CSSM and ARC Centre of Excellence for Dark Matter Particle Physics, Department of Physics, Adelaide University, Adelaide 5005, Australia}

\author{Giovani Dalla Valle Garcia}
\email{giovani.dallavallegarcia@unimelb.edu.au}
\affiliation{ARC Centre of Excellence for Dark Matter Particle Physics, School of Physics, University of Melbourne, Parkville, Victoria 3010, Australia}

\author{Xuan-Gong Wang}
\email{xuan-gong.wang@adelaide.edu.au}
\affiliation{CSSM and ARC Centre of Excellence for Dark Matter Particle Physics, Department of Physics, Adelaide University, Adelaide 5005, Australia}

\author{Anthony W. Thomas}
\email{anthony.thomas@adelaide.edu.au}
\affiliation{CSSM and ARC Centre of Excellence for Dark Matter Particle Physics, Department of Physics, Adelaide University, Adelaide 5005, Australia}

\author{Martin J. White}
\email{martin.white@adelaide.edu.au}
\affiliation{CSSM and ARC Centre of Excellence for Dark Matter Particle Physics, Department of Physics, Adelaide University, Adelaide 5005, Australia}

\begin{abstract}

The LUX–ZEPLIN (LZ) experiment has reported an inelastic nuclear-recoil candidate at $E_{\rm nr}=248\pm23_{\rm stat}\pm23_{\rm sys}\,{\rm keV}$.
We test whether endothermic pseudo-Dirac dark matter coupled to a kinetically mixed dark photon can account for this event while reproducing the observed relic abundance.
A five-parameter fit using an energy-only LZ likelihood and a relic-density likelihood favors TeV-scale dark matter with a mass splitting of a few hundred keV, and a GeV-scale mediator.
The recoil energy places the splitting near its kinematic ceiling, while secluded annihilation fixes the dark gauge coupling to a good approximation as a function of the dark-matter mass.
Representative points predict $\mathcal{O}(1)$ accepted LZ event candidate and the observed DM relic density.
Because the preferred splitting lies below the $e^+e^-$ threshold, the viability of the minimal model depends on the late-time excited-state abundance; a transition dipole can provide additional depletion, and the required light mediator remains testable in accelerator searches.

\end{abstract}

\maketitle

\section{Introduction}
\label{sec:introduction}

Recently, the LUX–ZEPLIN (LZ) collaboration recently reported one high-energy event consistent with a nuclear recoil at $E_{\rm nr}=248\pm23_{\rm stat}\pm23_{\rm sys}\,{\rm keV}$ in an exposure of $2.84\,{\rm tonne\,yr}$ \cite{LZ:2026highER}.
Across the tested dark matter (DM) effective operators, the maximum local significance is $3.4\sigma$, while the global significance after accounting for the look-elsewhere effect is $2.6\sigma$ \cite{LZ:2026highER}.
The unusually high recoil energy motivates an endothermic inelastic dark matter (iDM) interpretation, which we investigate here.

In iDM scenarios, two dark-sector states are separated by a small mass splitting $\delta$ \cite{Tucker-Smith:2001myb}.
Scattering into the heavier state requires extra kinetic energy and therefore a higher minimum incident velocity than elastic scattering.
This suppresses low-energy recoils and shifts the signal to higher recoil energies.
Recent interpretations of the LZ event include thermal Higgsino and broader BSM theories~\cite{Freese:2026sga, Fan:2026hig, Wu:2026nhi, Yin:2026pq, Visinelli:2026kgt, Smirnov:2026LZ, Du:2026LZ, Bose:2026ndd, Bisal:2026khf, Ahmed:2026kan, Lee:2026wof, Abdullahi:2023tyk, Lee:2026jxl, Lee:2026xxh, Yuan:2026djt, Cheung:2026byg, Du:2026lpa, Ahmed:2026qjg, Liang:2026coz, Yang:2026wpb, Wang:2026ytg, Okada:2026ypw}, endothermic dark-photon scattering near the kinematic limit \cite{Su:2026lzr,DiMauro:2026ldr,Yamashita:2026ump}, and exothermic dark matter \cite{Dent:2026bji, deLima:2026shq, Baer:2026fpy}.
Other studies consider a seesaw extension \cite{Das:2026uyy}, fermionic absorption \cite{Lou:2026idn}, singlino dark matter \cite{Chattopadhyay:2026LZ}, atmospheric-neutrino up-scattering \cite{Jeesun:2026LZ}, or an axion portal \cite{Unwin:2026LZ}.
Indirect searches, annual modulation, and the LZ high-energy sideband provide complementary tests \cite{Fan:2026hig, Wu:2026nhi, McCabe:2026LZ, Rodd:2026LZ, DiMauro:2026dqp}.

We study whether pseudo-Dirac DM coupled to a kinetically mixed dark photon can explain the LZ event while reproducing the observed relic abundance.
Dark photons provide a simple portal between the Standard Model (SM) and a hidden sector \cite{Fayet:1980ad,Fayet:1980rr,Holdom:1985ag,Okun:1982xi,Fabbrichesi:2020wbt,Filippi:2020kii}.
A small Majorana mass splits the dark fermion into two states, and the gauge interaction between the mass eigenstates is off diagonal\footnote{For a pseudo-Dirac model with a dark Higgs mediator see Ref. \cite{DallaValleGarcia:2024zva}, while a partial review on iDM can be found in Ref. \cite{DallaValleGarcia:2025giv}.}.
Unlike an effective-operator description, our model treats the dark photon as a propagating mediator and keeps its mass and kinetic mixing as explicit parameters.
These parameters determine both nuclear scattering and accelerator signals, allowing dark-photon searches to test the LZ interpretation independently.

We construct a likelihood from the reconstructed recoil energy of the LZ event and combine it with a relic-density likelihood for the five-parameter model.
We treat the dark photon as an explicit mediator and include coherent $A^\prime$ and $Z$ exchange in the scattering amplitude.
We also  discuss the late-time excited-state abundance and compare the preferred parameter region with accelerator constraints.

This paper is organized as follows.
Section~\ref{sec:model} defines the model.
Section~\ref{sec:rates} presents the endothermic kinematics, the xenon scattering rate, the relic-density calculation, and the LZ recoil-energy likelihood.
Section~\ref{sec:results} presents the results of the  fit.
Section~\ref{sec:removal} examines the late-time excited-state abundance.
Section~\ref{sec:constraints} discusses accelerator constraints, and Section~\ref{sec:conclusions} summarizes our conclusions.

\section{The model}
\label{sec:model}
 
The dark-photon model extends the SM gauge sector by an additional broken $U(1)^\prime$ gauge symmetry with associated gauge field $\hat X_\mu$.
The corresponding gauge-sector Lagrangian is \cite{Holdom:1985ag,Okun:1982xi}
\begin{equation}\label{eq:lagrangian}\begin{split}
	\mathcal{L}_{\mathrm{gauge}}\supset
&-\frac14\hat B_{\mu\nu}\hat B^{\mu\nu}
-\frac14\hat X_{\mu\nu}\hat X^{\mu\nu} \\
&+\frac{\epsilon}{2c_W}\hat X_{\mu\nu}\hat B^{\mu\nu}
+\frac12m_{X,0}^2\hat X_\mu\hat X^\mu ,
\end{split}\end{equation}
Here $\hat B_\mu$ is the hypercharge gauge field, $c_W\equiv\cos\theta_W$, and $\epsilon$ is the kinetic-mixing parameter.
In this setup, the kinetic-mixing term provides the portal between the SM and $U(1)^\prime$ sectors, while the mass eigenstate predominantly aligned with $\hat X_\mu$ is identified as the dark photon after diagonalization.
The vector mass is represented by an effective Proca term $m_{X,0}$, without specifying the associated symmetry-breaking sector.

The mixed kinetic terms of the gauge fields $\hat B_\mu$ and $\hat X_\mu$ in Eq. (1) are diagonalized and canonically normalized by expressing them in terms of the redefined fields $B_\mu$ and $X_\mu$ \cite{Kribs:2020vyk}
\begin{equation}
	\hat B_\mu=B_\mu+\frac{\epsilon}{c_W\kappa}X_\mu,
\qquad
\hat X_\mu=\frac{X_\mu}{\kappa},
\end{equation}
where $\kappa\equiv\sqrt{1-\epsilon^2/c_W^2}$.
After electroweak symmetry breaking, the photon and the unmixed SM-like $Z$ state, ${\bar{Z}}$, are defined as
\begin{equation}
	A_\mu=s_WW^3_\mu+c_WB_\mu,
\qquad
\bar Z_\mu=c_WW^3_\mu-s_WB_\mu,
\end{equation}
where $s_W\equiv\sin\theta_W$ and $\theta_W$ is the weak mixing (Weinberg) angle.
In the basis $(\bar Z_\mu,X_\mu)$, the neutral-vector mass matrix is
\begin{equation}\begin{split}
M_V^2=m_{\bar Z}^2
\begin{pmatrix}
1&-\epsilon_W\\
-\epsilon_W&\epsilon_W^2+\rho^2
\end{pmatrix}, \\
\epsilon_W\equiv\frac{\epsilon\tan\theta_W}{\kappa},
\quad
\rho\equiv\frac{m_{X,0}}{m_{\bar Z}\kappa}.
\end{split}\end{equation}
The physical $(Z, A^\prime)$ mass eigenvalues are then
\begin{small}
\begin{equation}
	m_{Z,A^\prime}^2
=\frac{m_{\bar Z}^2}{2}
\left[
1+\epsilon_W^2+\rho^2
\pm\sqrt{\left(1+\epsilon_W^2+\rho^2\right)^2-4\rho^2}
\right],
\end{equation}
\end{small}
For $m_{A^\prime}<m_Z$, the upper and lower signs correspond to $m_Z^2$ and $m_{A^\prime}^2$, respectively.

The dark sector contains a Dirac fermion $\psi$ of unit $U(1)^\prime$ charge, with covariant derivative $D_\mu=\partial_\mu+i g_D\hat X_\mu$ and a small $U(1)^\prime$-breaking Majorana mass $m_M\ll m_D$, described by the Lagrangian 
\begin{equation}
	\mathcal{L}_{\psi}
=\bar\psi i\gamma^\mu D_\mu\psi
-m_D\bar\psi\psi
-\frac{m_M}{2}\left(\bar\psi^c\psi+\mathrm{h.c.}\right).
\end{equation}
The Majorana mass term $m_M$ splits the Dirac fermion $\psi$ into two Majorana mass eigenstates, $\chi_1$ and $\chi_2$.
Using the phase convention
\begin{equation}
\psi=\frac{1}{\sqrt2}\left(\chi_1+i\chi_2\right),
\quad
\chi_i=\chi_i^c,	
\end{equation}
the corresponding mass eigenvalues are
\begin{equation}
m_{1,2}=m_D\mp m_M.
\end{equation}

Since the vector bilinears of individual Majorana fields vanish, $\bar\chi_i\gamma^\mu\chi_i=0$, the dark current is purely off diagonal.
The interaction can be written as
\begin{equation}\label{eq:la-XVint}
\begin{split}
\mathcal{L}_{\chi V}
&=g_D\hat X_\mu\bar\psi\gamma^\mu\psi\\
&=\frac{i g_D}{\kappa}X_\mu\bar\chi_1\gamma^\mu\chi_2\\
&=i\left(g_{\chi A^\prime}A^\prime_\mu+g_{\chi Z}Z_\mu\right)
\bar\chi_1\gamma^\mu\chi_2,
\end{split}\end{equation}
where the couplings to the physical vector bosons are
\begin{equation}
	g_{\chi A^\prime}=\frac{g_Dc_\alpha}{\sqrt{1-\epsilon^2/c_W^2}},
\quad
g_{\chi Z}=\frac{g_Ds_\alpha}{\sqrt{1-\epsilon^2/c_W^2}} .
\end{equation}
Here $c_\alpha\equiv\cos\alpha$ and $s_\alpha\equiv\sin\alpha$, where $\alpha$ is the $\bar Z$–$X$ mixing angle that diagonalizes the neutral-vector mass matrix $M_{V}^2$; the convention is chosen such that $A^\prime_\mu\to X_\mu$ as $\epsilon\to0$.
The corresponding couplings to SM fermions are given in Ref. \cite{Kribs:2020vyk}.
Loop-induced elastic interactions are   neglected \cite{Bell:2018zra}.

In summary, the model is specified by five input parameters,
\begin{equation}\label{eq:model-params}
	m_\chi\equiv m_1,\quad
\delta\equiv m_2-m_1,\quad
m_{A^\prime},\quad\epsilon,\quad g_D
.
\end{equation}


\section{Inelastic scattering, relic abundance, and the LZ likelihood}
\label{sec:rates}
This section combines endothermic kinematics, xenon scattering, two-state freeze-out, and a simplified LZ detector model to construct the energy-only likelihood used in our scan. Numerical illustrations use the three benchmarks in Table~\ref{tab:benchmarks}, with masses near $500\,{\rm GeV}$, $1\,{\rm TeV}$, and $3\,{\rm TeV}$, each yielding $\mathcal{O}(1)$ accepted event and the observed relic abundance. The full scan and excited-state evolution are presented in Sec~
\ref{sec:results} and Sec~\ref{sec:removal}.

\begin{table}[htbp]
\centering
\caption{Benchmark parameters and derived quantities. Kinematic and
direct-detection quantities are evaluated for ${}^{131}\mathrm{Xe}$ at
$E_R=248\,\mathrm{keV}$.}
\label{tab:benchmarks}
\begin{tabular}{lccc}
\toprule
Parameter & \texttt{BM-A} & \texttt{BM-B} & \texttt{BM-C} \\
\midrule
$m_\chi$ [GeV]
  & $500.1$ & $1024.7$ & $2867.8$ \\
$\delta$ [keV]
  & $326.6$ & $357.4$ & $376.2$ \\
$m_{A^\prime}$ [GeV]
  & $1.030$ & $1.066$ & $1.082$ \\
$\epsilon$
  & $0.0605$ & $0.0380$ & $0.0222$ \\
$g_D$
  & $0.468$ & $0.664$ & $1.112$ \\
\midrule
\multicolumn{4}{l}{{Derived}} \\
\midrule
\addlinespace
$\delta/\delta_{\max}$
  & $0.944$ & $0.948$ & $0.950$ \\
$v_\delta$ [km\,s$^{-1}$]
  & $774$ & $768$ & $760$ \\
$E_R^{\mathrm{peak}}$ [keV]
  & $239$ & $334$ & $364$ \\
$\sigma_{\chi p}^{\mathrm{SI}}$ [pb]
  & $7.67\times10^{3}$ & $5.30\times10^{3}$ & $4.80\times10^{3}$ \\
$\mu_s$ [events]
  & $1.143$ & $1.074$ & $1.161$ \\
$f_s(248\,\mathrm{keV})$ [keV$^{-1}$]
  & $1.45\times10^{-2}$ & $1.59\times10^{-2}$ & $1.49\times10^{-2}$ \\
$\Omega h^2$
  & $0.1150$ & $0.1204$ & $0.1192$ \\
$\ln\mathcal{L}_{\mathrm{LZ}}$
  & $-5.250$ & $-5.150$ & $-5.225$ \\
\midrule
\multicolumn{4}{l}{{Excited state}} \\
\midrule
\addlinespace
$T_{\mathrm{kd}}$ [MeV]
  & $0.81$ & $1.04$ & $1.35$ \\
$f_2^{\mathrm{conv}}$
  & $2.44\times10^{-3}$ & $9.42\times10^{-4}$ & $2.77\times10^{-4}$ \\
$\tau_{3\gamma}$ [s]
  & $6.30\times10^{16}$ & $2.75\times10^{16}$ & $1.54\times10^{16}$ \\
$f_2(t_0)$
  & $2.51\times10^{-6}$ & $1.55\times10^{-10}$ & $2.65\times10^{-16}$ \\
$\mu_s^{\mathrm{exo}}(t_0)$ [events]
  & $3.44\times10^{6}$ & $46.6$ & $2.00\times10^{-5}$ \\
\bottomrule
\end{tabular}
\end{table}

\subsection{LZ candidate and endothermic kinematics}
\label{sec:lzkine}
LZ reports a high-energy nuclear-recoil candidate consistent with an inelastic dark-matter scattering interpretation, with a reconstructed xenon recoil energy $E_{\rm nr}=248\pm23_{\rm stat}\pm23_{\rm sys}\,{\rm keV}$. In our model, this comes from the endothermic process $\chi_1+A\longrightarrow\chi_2+A.$
Only the fast tail of the local-halo speed distribution in the laboratory frame can contribute. For the kinematic estimates below, we use the central reported energy as the reference recoil energy; the distinction between true and reconstructed recoil energy, together with detector smearing, is restored in Sec.~\ref{sec:loglLZ}.

For a target isotope of mass $m_A$, corresponding to the DM-isotope reduced mass $\mu_{\chi A}$, the minimum incoming speed required to produce a true recoil energy $E_R$ is
\begin{equation}
	v_{\min}^A(E_R) =\sqrt{\frac{m_AE_R}{2\mu_{\chi A}^2}} +\frac{\delta}{\sqrt{2m_AE_R}}.
\end{equation}
The second term is the additional kinematic cost of producing the $\chi_2$ particle. Unlike elastic scattering, $v_{\rm min}^A$ decreases with $E_R$, so endothermic scattering can favor recoils of several hundred keV while suppressing low-energy events.

For the fiducial halo model, we use the tabulated Eddington-inverted speed distribution of the tapered Einasto profile in Ref. \cite{Wu:2026nhi}, with $\rho_\chi=0.4\,{\rm GeV\,cm^{-3}}$, $v_{\rm esc}=544\,{\rm km\,s^{-1}}$, and $v_{\rm Earth}=254\,{\rm km\,s^{-1}}$. The laboratory-frame distribution then has finite support up to $v_{\max}=798\,{\rm km\,s^{-1}}$. We neglect annual modulation and fix $v_{\rm Earth}$ at this representative value. That is adequate for the kinematic bound, but a seasonal variation of order $30\,{\rm km\,s^{-1}}$ is comparable to the remaining margin near the endpoint and should be included in a dedicated analysis.

Requiring the candidate recoil to be kinematically accessible, $v_{\min}^A(E_R)\leq v_{\max}$, gives an upper bound on the mass splitting,
\begin{equation}
	\delta_{\max}(m_\chi) = \sqrt{2m_AE_R} 
	\left[ v_{\max} - \sqrt{\frac{m_AE_R}{2\mu_{\chi A}^2}} \right].
\end{equation}
For $^{131}{\rm Xe}$ at $E_R=248\,{\rm keV}$, this becomes
\begin{equation}\label{eq:deltamax}
	\delta_{\max} = 406.6\,{\rm keV} \left(1-\frac{74.4\,{\rm GeV}}{m_\chi}\right).
\end{equation}
Thus, at a true recoil energy of exactly $248\,{\rm keV}$, elastic scattering from $^{131}{\rm Xe}$ is kinematically forbidden for $m_\chi<74.4\,{\rm GeV}$, while the maximum allowed endothermic splitting approaches $406.6\,{\rm keV}$ in the heavy-DM limit. We use Eq.~(\ref{eq:deltamax}) as an approximate energy cut in the parameter scan, rather than treating it as an exact detector sensitivity threshold for all isotopes. Because resolution smearing allows events reconstructed at $248\,{\rm keV}$ to originate from nearby true recoil energies, heavier xenon isotopes permit a slightly larger asymptotic splitting (about $419\,{\rm keV}$ for $^{136}{\rm Xe}$).

The same kinematics determine the range of recoil energies available at a fixed incoming speed. The allowed interval is
\begin{equation}
	E_R^\pm(v)
=
\frac{\mu_{\chi A}^2}{2m_A}
\left(v\pm\sqrt{v^2-v_\delta^2}\right)^2,
\quad
v_\delta=\sqrt{\frac{2\delta}{\mu_{\chi A}}} .
\end{equation}
No scattering is possible for $v<v_\delta$. At threshold, $v=v_\delta$, the interval collapses to
\begin{equation}
	E_R^\ast = \frac{\mu_{\chi A}\delta}{m_A} .
\end{equation}
The recoil energy $E_R^\ast$ is the point at which $v_{\min}(E_R)$ is minimized. 
For $^{131}$Xe at a recoil energy of 248 keV, the minimum incoming speed required for up-scattering is about 774 km/s. The benchmark BM-B has $m_\chi=1025\,{\rm GeV}$ and $\delta=357\,{\rm keV}$, corresponding to $\delta/\delta_{\max}=0.95$. On $^{131}$Xe this gives an inelastic threshold speed of about 768 km/s. Both speeds therefore lie within about $4\%$ of the laboratory-frame cutoff $v_{\max}$.

\subsection{Xenon scattering rate}
\label{sec:xenonsr}
Because the diagonal Majorana vector current vanishes in Eq.~(\ref{eq:la-XVint}), the detection signal results from the inelastic process $\chi_1 N \to \chi_2 N$, with coherent $A^\prime$ and $Z$ exchange.

The quark-level vector current is matched onto separate proton and neutron amplitudes using the conserved $u$- and $d$-quark vector charges. The point-nucleon couplings are
\begin{equation}
	g_{Vp}=2g_{Vu}+g_{Vd},
\qquad
g_{Vn}=g_{Vu}+2g_{Vd}.
\end{equation}
Here $g_{VN}$ denotes the effective vector coupling of nucleon $N$. 

The corresponding point-nucleon operator coefficients are
\begin{equation}\label{eq:cn}
	C_N(q) = \frac{g_{\chi A^\prime}g_{A^\prime N}}{q^2+m_{A^\prime}^2}
+ \frac{g_{\chi Z}g_{Z N}}{q^2+m_Z^2}.
\end{equation}
Here $g_{A^\prime N}$ and $g_{Z N}$ denote the effective vector couplings of the dark photon and the $Z$ boson to the nucleon vector current. 
Finite nucleon size is included through the electric form factors, using the dipole parametrization for the proton and the Galster parametrization for the neutron \cite{Galster:1971kv, Ye:2017gyb},
\begin{equation}\begin{split}
G_E^{\rm p}(q^2) &= \left(1+\frac{q^2}{0.71\,{\rm GeV}^2}\right)^{-2}, \\
G_E^{\rm n}(q^2) &= \frac{1.913\,\tau}{1+5.6\tau}\,G_E^{\rm p}(q^2), \quad 
\tau =\frac{q^2}{4m_N^2}.	
\end{split}\end{equation}
The finite-momentum transition amplitudes are then
\begin{equation}\label{eq:dmdd-amp}\begin{split}
\mathcal A_{\rm p}^{\rm SI}(q) = \frac12\left[C_{\rm p}(q)G_E^{\rm p}(q^2)+C_{\rm n}(q)G_E^{\rm n}(q^2)\right], \\
\mathcal A_{\rm n}^{\rm SI}(q) = \frac12\left[C_{\rm n}(q)G_E^{\rm p}(q^2)+C_{\rm p}(q)G_E^{\rm n}(q^2)\right].
\end{split}\end{equation}
At vanishing momentum transfer $G_E^p$ is 1, $G_E^n$ vanishes, and both mediators give a vanishing $C_{\rm n}$. The interaction is therefore proton-only. The corresponding reference cross section is
\begin{equation}\label{eq:dmdd-xsect}
	\sigma_{\chi N}^{\rm SI}
\equiv \frac{4\mu_{\chi N}^2}{\pi}
\left|\mathcal{A}_N^{\rm SI}(0)\right|^2,
\quad
\mu_{\chi N} = \frac{m_\chi m_N}{m_\chi+m_N} .
\end{equation}
Equation~(\ref{eq:dmdd-xsect}) describes an inelastic transition and cannot be compared directly with elastic exclusion curves. At 248 keV on $^{131}{\rm Xe}$ the momentum transfer is about $\sqrt{2m_AE_R}\simeq0.246\,{\rm GeV}$.  $G_E^{\rm p}$ and $G_E^{\rm n}$ are then 0.849 and 0.0254, and for the benchmark couplings they reduce the coherent xenon response by a factor of about 0.784 relative to point nucleons. The rate is therefore evaluated with Eq.~(\ref{eq:dmdd-amp}), not with the zero-momentum reference.

The dark current is purely vector, so there is no axial–axial spin-dependent operator. A residual vector–axial term is velocity suppressed and receives no coherent nuclear enhancement. At these kinematics the effective proton-level ratio is $2.6\times10^{-5}$ relative to the spin-independent response, and its contribution to the $^{131}{\rm Xe}$ rate is below $10^{-8}$ of the coherent response. We therefore retain only the spin-independent mass response. Loop-induced elastic scattering is  neglected here~\cite{Bell:2018zra}; a residual $\chi_2$ population is treated in Sec.~\ref{sec:removal}.

The nucleon amplitudes determine the spin-independent nuclear response
\begin{small}\begin{equation}
	\mathcal{R}_A(q)
=
\frac{4\pi}{2J_A+1}
\begin{pmatrix}c_0 (q)&c_1 (q)\end{pmatrix}
\begin{pmatrix}
W_{00}^A(q)&W_{01}^A(q)\\
W_{01}^A(q)&W_{11}^A(q)
\end{pmatrix}
\begin{pmatrix}c_0(q)\\c_1(q)\end{pmatrix},
\end{equation}
\end{small}
where $c_0$ is the isoscalar combination $\mathcal{A}_{\rm p}^{\rm SI}+\mathcal{A}_{\rm n}^{\rm SI}$, and $c_1$ is the isovector combination $\mathcal{A}_{\rm p}^{\rm SI}-\mathcal{A}_{\rm n}^{\rm SI}$. At zero momentum transfer this reduces to $Z_A^2|\mathcal A_p^{\rm SI}|^2$.

The corresponding differential cross section on isotope $A$ is
\begin{equation}
	\frac{{\rm d}\sigma_A}{{\rm d}E_R}
=
\frac{2m_A}{\pi v^2}\,
\mathcal R_A(q)\,
\Theta\!\left(v-v_{\min}^A(E_R)\right).
\end{equation}
The step function $\Theta$ restricts scattering to incoming speeds above $v_{\min}^A$.
The differential event rate is
\begin{equation}
	\frac{{\rm d}R}{{\rm d}E_R}
=
\frac{2\rho_\chi}{\pi m_\chi}
\sum_A\xi_A\,
\mathcal{R}_A(q_A)\,
\eta\!\left(v_{\min}^A(E_R)\right),
\end{equation}
where $\xi_A$ is the xenon mass fraction of each isotope, $q_A$ is the corresponding momentum transfer, and $\eta$ is the mean inverse speed of halo particles above threshold:
\begin{equation}
	\eta(v_{\min})
= \int_{v_{\min}}^{v_{\max}}
\frac{f_{\rm lab}(v)}{v}\,{\rm d}v .
\end{equation}
We use the $W_M$ responses of WimPyDD \cite{Jeong:2021wimpydd} for $A=128$-$136$ and a Helm form factor for $^{124}{\rm Xe}$ and $^{126}{\rm Xe}$.

When the mediator is heavy compared with the momentum transfer, the amplitude scales as $\epsilon g_D/m_{A^\prime}^2$ and the rate as $m_{A^\prime}^{-4}$ at fixed $\epsilon$ and $g_D$. Reducing the mediator mass by three decades therefore enhances the rate by twelve orders of magnitude. This is only a limiting estimate. We checked that, at the event rate required by LZ, the preferred $m_{A^\prime}$ interval is around $1\,{\rm GeV}$, comparable to the momentum transfer, so the rate uses the full propagator in Eq.~(\ref{eq:cn}).

\subsection{Thermal relic abundance}
\label{sec:relic}
The above discussion fixes the region favored by the LZ candidate: $m_\chi$ above $74.4\,{\rm GeV}$, $\delta$ of at most a few hundred keV, and $m_{A^\prime}$ around $1\,{\rm GeV}$. In this work we adopt the thermal WIMP freeze-out mechanism. The corresponding freeze-out therefore occurs in the secluded regime $m_{A^\prime}\ll m_\chi$ \cite{Pospelov:2008zw}, so the dominant annihilation channels are $\chi_1\chi_1\to A^\prime A^\prime$ and $\chi_2\chi_2\to A^\prime A^\prime$, through $t$- and $u$-channel exchange of the other dark state. The mixed process $\chi_1\chi_2\to{\rm SM}$ proceeds through $s$-channel $A^\prime$ and $Z$ exchange and is suppressed by $\epsilon$.

The ratio $\delta/m_\chi$ is at most a few times $10^{-6}$, so the splitting is negligible at $T_f\sim m_\chi/27$. Both Majorana states remain populated and must be included together in the thermal calculation \cite{Griest:1990kh,Gondolo:1990dk}. We compute the relic abundances of both dark-matter particles simultaneously. Equal population of the two states gives the effective annihilation rate
\begin{equation}\label{eq:dm-ann}
	\langle\sigma_{\rm eff}v\rangle
\simeq \frac14\left( \langle\sigma_{11}v\rangle + \langle\sigma_{22}v\rangle + 2\langle\sigma_{12}v\rangle
\right).
\end{equation}
The two equal-state channels each contribute about half of the total rate, with all other channels well below $1\%$. In the secluded limit this rate is
\begin{equation}\label{eq:sigmav}
	\langle\sigma v\rangle_{\chi\chi\to A^\prime A^\prime}
\simeq
\frac{g_D^4}{16\pi m_\chi^2} .
\end{equation}
We checked that Eq.~(\ref{eq:sigmav}) reproduces the numerical freeze-out rate to within $0.9\%$ and follows the expected $g_D^4$ scaling.

The dark-photon mass is introduced as a Proca term in Eq.~(\ref{eq:lagrangian}). The tree-level $\chi\chi\to A^\prime A^\prime$ amplitude computed by CalcHEP can then grow unphysically through the longitudinal mode. The numerical analysis in Sec. 4 therefore imposes a restriction that keeps this growth under control. That restriction controls only the longitudinal mode. Sommerfeld and bound-state corrections are not included. With the relic $g_D$ and $m_{A^\prime}$ around $1\,{\rm GeV}$, $\alpha_D m_\chi/m_{A^\prime}$ remains large, so the $\Omega h^2$ and the relation between $g_D$ and $m_\chi$ obtained here are only tree-level freeze-out results.

In this minimal phenomenological model, the relic density therefore fixes $g_D$ as a function of $m_\chi$, whereas $\epsilon$ and $m_{A^\prime}$ control the direct-detection rate.

\subsection{LZ detector response and likelihood}
\label{sec:loglLZ}


The rate in Eq.~(\ref{eq:dm-ann}) is a function of the true recoil energy $E_R$. We construct an approximate spectrum in the reconstructed energy $E$ by folding the true recoil spectrum with the central efficiency curve of Fig. S2 of Ref. \cite{LZ:2026highER} and a Gaussian response\footnote{LZ has not released the full two-dimensional $(S1_c,\log_{10}S2_c)$ likelihood for the high-energy search. We therefore use the reconstructed-energy likelihood introduced in Ref.~\cite{Wu:2026nhi}.}. We take $\sigma_E=23\,{\rm keV}$ as a local proxy for the response width near $E_{\rm obs}=248\,{\rm keV}$, motivated by the reported statistical uncertainty on the reconstructed recoil energy of the candidate.

We adopt the following detector and background settings for the LZ high-energy nuclear-recoil search. The exposure $\mathcal E$ is $2.84\,{\rm tonne\,yr}$, or $1.037\times10^6\,{\rm kg\,day}$. The reconstructed-energy window runs from $5$ to $300\,{\rm keV}$, and the candidate energy $E_{\rm obs}$ is $248\,{\rm keV}$. The acceptance is approximately $0.96$ below $240\,{\rm keV}$, decreases to $0.95$ at the candidate energy, $0.49$ at $270\,{\rm keV}$, and $0.16$ at $280\,{\rm keV}$, and vanishes by $299\,{\rm keV}$. The expected background $\bar b$ is $0.0106$ events, with uncertainty $\sigma_b$ of $0.0008$, and the background energy density $f_b$ is taken flat across the window at $1/(295\,{\rm keV})$.

We model the energy response with the normalized Gaussian
\begin{equation}
	G(E\mid E_R,\sigma_E)
=
\frac{1}{\sqrt{2\pi}\sigma_E}
\exp\!\left(-\frac{(E-E_R)^2}{2\sigma_E^2}\right).
\end{equation}
The accepted reconstructed-energy spectrum is
\begin{equation}
	\frac{{\rm d}N_{\rm rec}}{{\rm d}E} = \mathcal E
\int {\rm d}E_R\, \frac{{\rm d}R}{{\rm d}E_R} \varepsilon_{\rm det}(E_R) G(E\mid E_R,\sigma_E).
\end{equation}
For our model, the true-recoil spectrum that contributes to this integral lies between $95$ and $442\,{\rm keV}$.
The expected signal yield in the reconstructed-energy window and the corresponding normalized distribution are then
\begin{equation} \begin{split}
&\mu_s(\theta)
=
\int_{5\,{\rm keV}}^{300\,{\rm keV}}
{\rm d}E\,
\frac{{\rm d}N_{\rm rec}}{{\rm d}E},
\\ 
&f_s(E\mid\theta)
=
\frac{1}{\mu_s(\theta)}
\frac{{\rm d}N_{\rm rec}}{{\rm d}E},
\\
\int_{5\,{\rm keV}}^{300\,{\rm keV}}
&f_s(E\mid\theta)\,{\rm d}E
=1 .
\end{split}\end{equation}
For one observed event, the energy-only likelihood is
\begin{equation}
	\begin{split}
\mathcal{L}(\theta,\mu_b)
= {}&
e^{-[\mu_s(\theta)+\mu_b]}
\big(
\mu_s(\theta)f_s(E_{\rm obs}\mid\theta)
+
\mu_bf_b
\big)\\
&\times
\exp\left[-\frac{(\mu_b-\bar b)^2}{2\sigma_b^2}\right],
\qquad
\mu_b\geq0 ,
\end{split}
\end{equation}
The Gaussian constraint makes $\mu_b$ a nuisance parameter, and the likelihood used in the scan is obtained by profiling,
\begin{equation}\label{eq:logLLZ}
\ln\mathcal{L}_{\rm LZ}(\theta)
= \max_{\mu_b\geq0} \ln\mathcal{L}(\theta,\mu_b) .	
\end{equation}
The fit depends on both the accepted yield and the reconstructed shape at the candidate energy. With a single event, the Poisson factor favours a yield of order one, but the event is also weighted by the local densities $\mu_s f_s(E_{\rm obs})$ and $\mu_b f_b$. A rate spread across the analysis window is therefore hard to distinguish from background, whereas a spectrum peaked near $248\,{\rm keV}$ is more distinctive. The background-only limit gives $\ln\mathcal{L}_{\rm LZ}=-10.242$.

Table~{\ref{tab:benchmarks}} gives the input parameters and derived quantities for the three benchmarks introduced above. Despite their distinct true recoil spectra, their values of $\ln\mathcal{L}_{\rm LZ}$ differ by only $0.10$. 

\begin{figure}
  \centering
  \includegraphics[width=\linewidth]{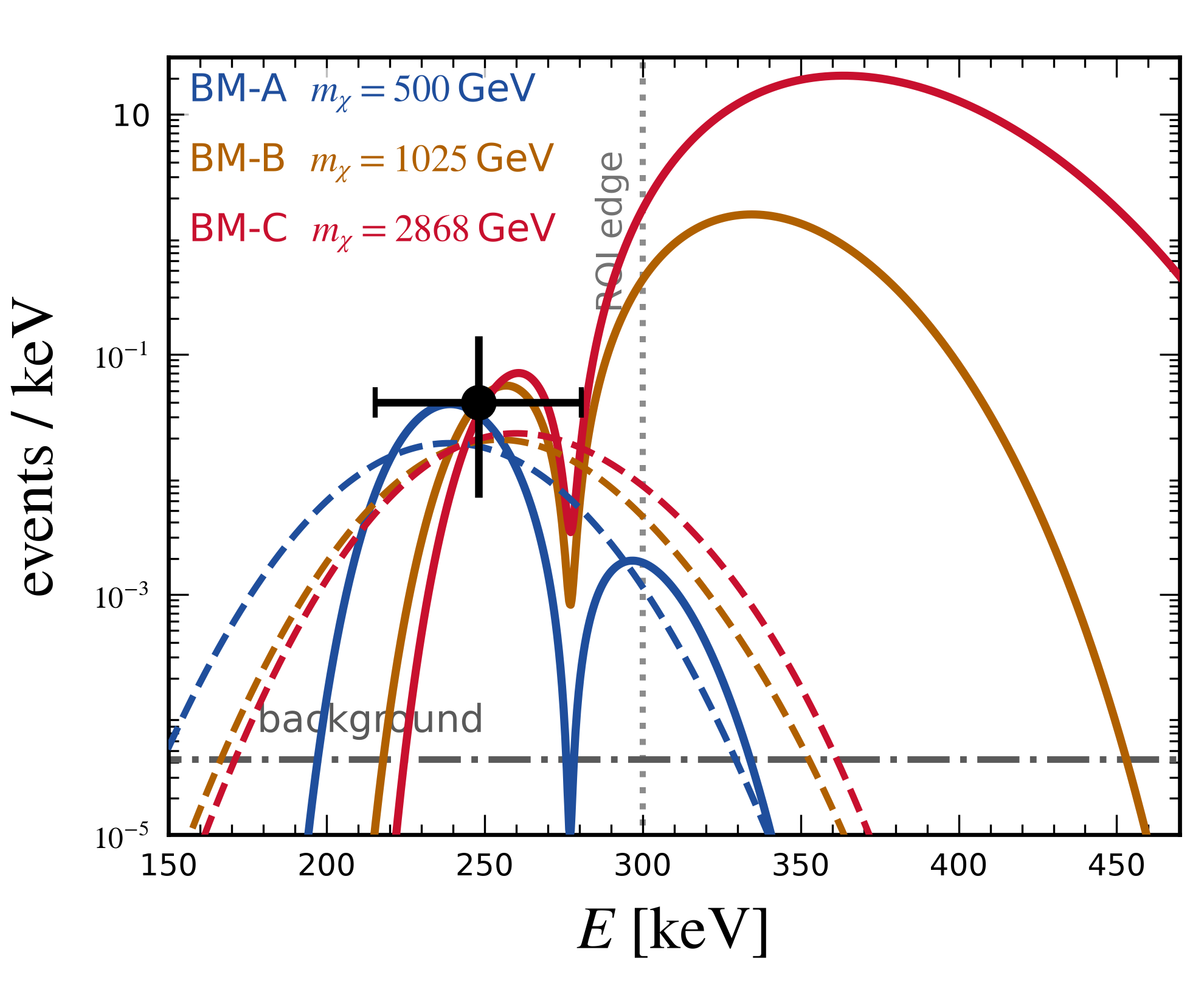}
  \caption{Recoil spectra of the three benchmarks in Table~\ref{tab:benchmarks}, multiplied by the exposure so that areas are event numbers. Solid curves: true recoil energy. Dashed curves: accepted reconstructed spectra $\mu_s f_s(E)$ after efficiency and $\sigma_E=23\,{\rm keV}$ resolution. Dot-dashed: constrained-background density $\bar b f_b=3.59\times10^{-5}\,{\rm events\,keV^{-1}}$. The dotted line is the upper edge of the analysis window. The point is the LZ candidate, with the reported energy uncertainty horizontally and the $68\%$ Poisson interval for one event vertically. }
  \label{fig:spectrum}
\end{figure}

Figure 1 shows that the three benchmarks produce similar reconstructed signals despite different true spectra. \texttt{BM-A} peaks at $239\,{\rm keV}$, within the accepted range, while \texttt{BM-B} and \texttt{BM-C} peak at $334\,{\rm keV}$ and $364\,{\rm keV}$, both above the range with nonzero acceptance. After the application of the detector response, the reconstructed shapes at the candidate energy agree to within $10\%$ and the accepted yields to within $8\%$, so an energy-only analysis of one event cannot distinguish these cases.

For \texttt{BM-B} the true spectrum peaks at $E_R^{\rm peak}\simeq334\,{\rm keV}$, where the efficiency has already vanished, and only the low-energy shoulder is visible. Of the $91$ recoils produced in the full exposure, $1.20$ enter the accepted yield, an accepted fraction of $1.3\%$. The present analysis is therefore sensitive only to the shoulder of a spectrum whose maximum lies above the current selection. The true spectra show a nuclear diffraction feature of the isotope-summed $W_M$ response near $275\,{\rm keV}$. The detector reconstruction smears this structure away.

\section{Numerical Results}
\label{sec:results}

\subsection{Setup}
\label{sec:scan-setup}

We sample the five parameters of Eq.~(\ref{eq:model-params}) with an affine-invariant ensemble Markov chain Monte Carlo \cite{Goodman:2010dyf,Foreman-Mackey:2012any} implemented in Jarvis-HEP \cite{Guo:2026kfy}. The combined priors are
\begin{equation}\label{eq:priors}
\begin{aligned}
200 {\rm GeV} \leq &m_\chi \leq 5 {\rm TeV},\\
100 {\rm keV} \leq &\delta \leq 600 {\rm keV},\\
0.1 {\rm GeV}\leq &m_{A^\prime} \leq 80 {\rm GeV}, \\
10^{-9} \leq &\epsilon \leq 0.5, \\
0.1 \leq &g_D \leq 2.9,
\end{aligned}
\end{equation}
all log-uniform.
The likelihood used in this scan is defined as
\begin{equation}\begin{split}
\mathcal{L}_{\rm tot} &= \mathcal{L}_{\rm LZ} \times \mathcal{L}_\Omega, \\
\ln\mathcal{L}_\Omega &=-\frac12
\left(\frac{\Omega h^2-0.1200}{0.01206}\right)^2, 
\end{split}
\end{equation}
where $\mathcal{L}_{\rm LZ}$ is the profiled energy-only proxy of Eq.~(\ref{eq:logLLZ}). The width $0.01206$ is the Planck uncertainty $0.0012$ \cite{Planck:2018vyg,ParticleDataGroup:2024cfk} added in quadrature with a $10\%$ theoretical error on $\Omega h^2=0.1200$. 

The model is implemented in FeynRules \cite{Christensen:2009jx,Alloul:2013bka}. The phenomenology of Sec.~\ref{sec:rates} is incorporated into micrOMEGAs-7 \cite{Belanger:2026asz,Alguero:2023zol}, which calls CalcHEP \cite{Belyaev:2012qa} internally for tree-level matrix elements.

Because of the Proca mass term discussed in Sec.~\ref{sec:relic}, we impose a hard cut on the sampled parameters,
\begin{equation}\label{eq:hardcut}
	\frac{\delta\,m_\chi}{m_{A^\prime}^{2}} < 1.33,
\end{equation}
so that the unphysical enhancement of the tree-level rate remains below about $10\%$. This restriction comes from limitations of the present model and of the numerical calculation, not from a physical bound.

\subsection{Results}
\label{sec:preferred-region}

\begin{figure*}
  \centering
  \includegraphics[width=0.45\linewidth]{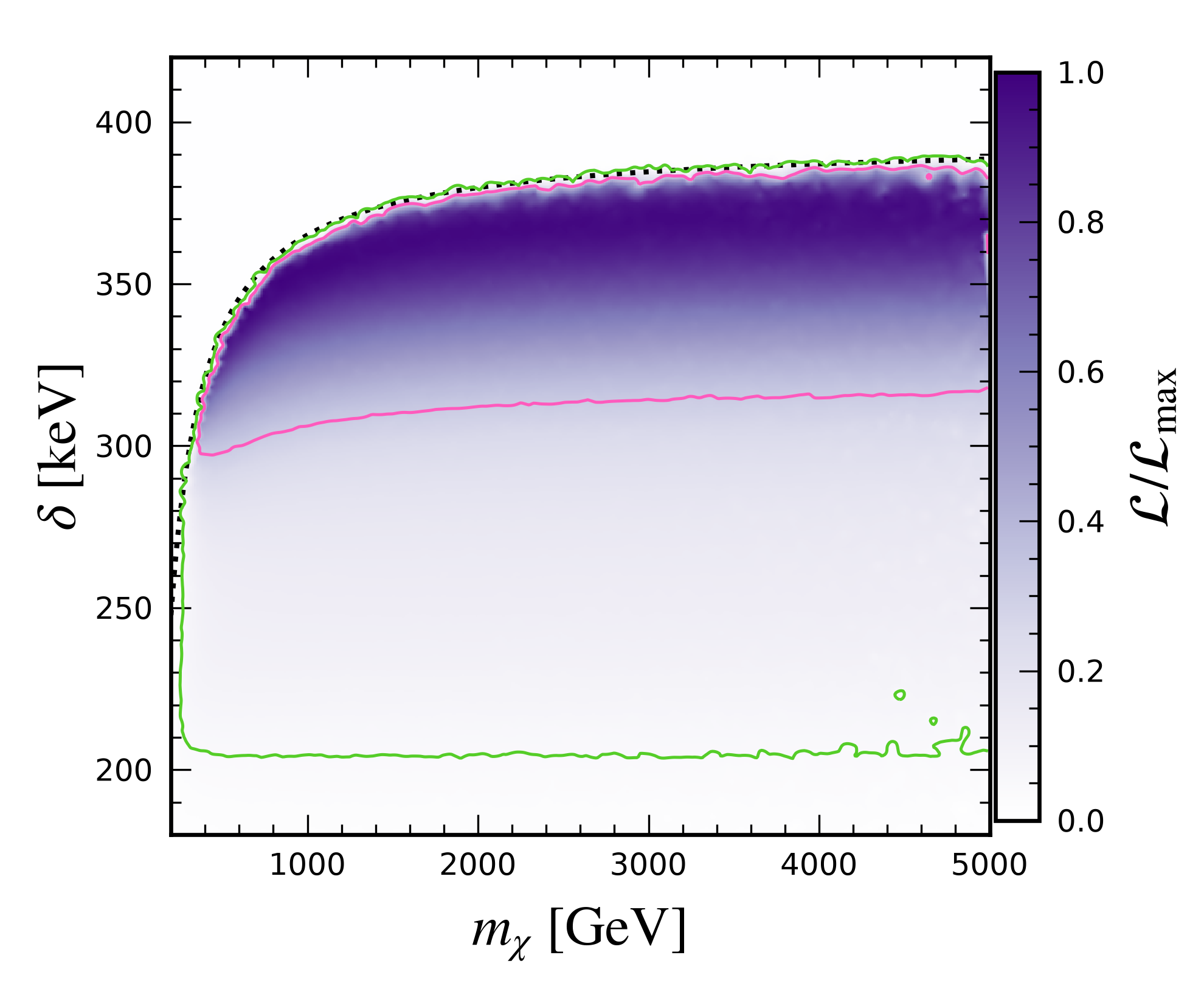}
  \includegraphics[width=0.45\linewidth]{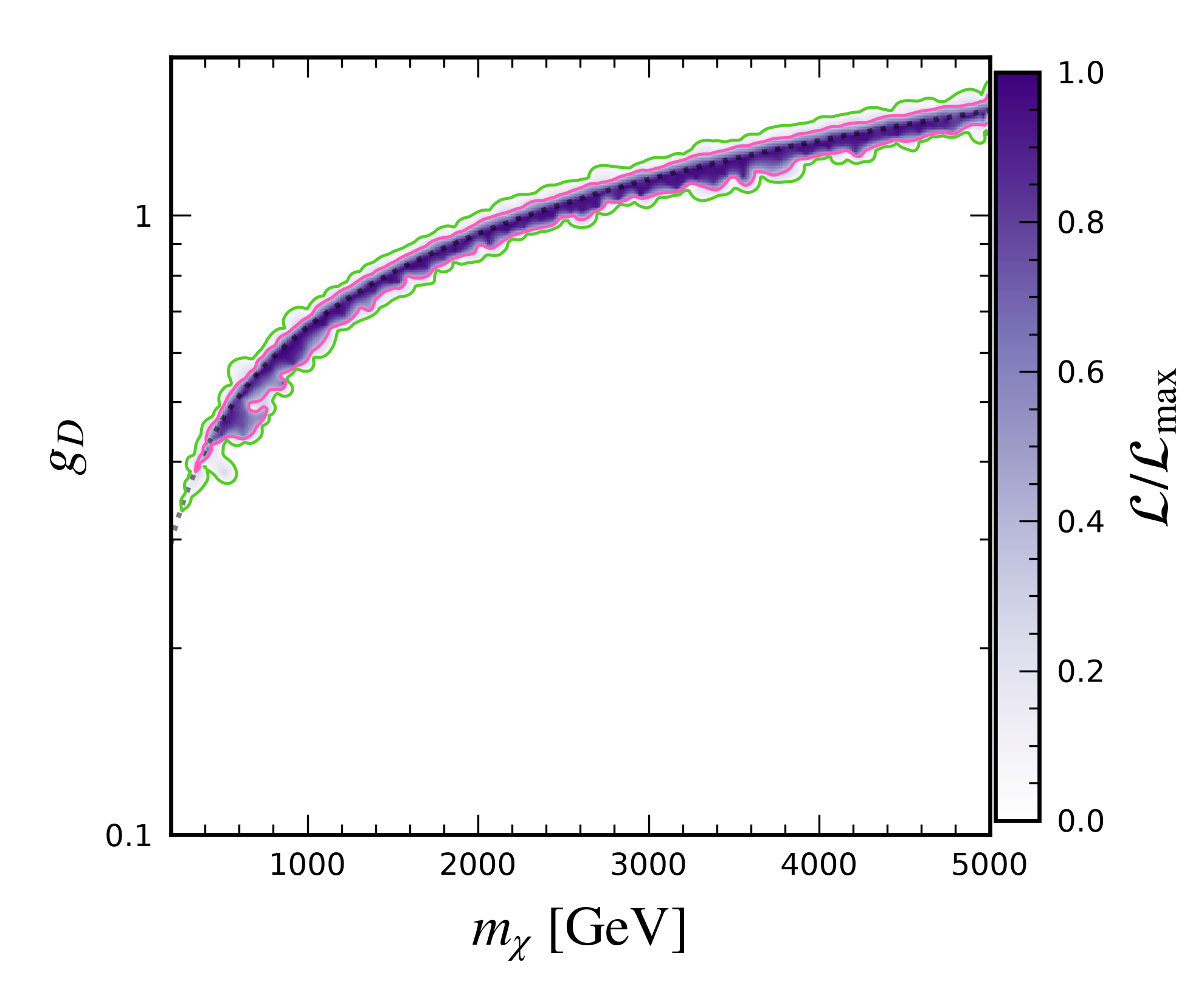} \\
  \includegraphics[width=0.45\linewidth]{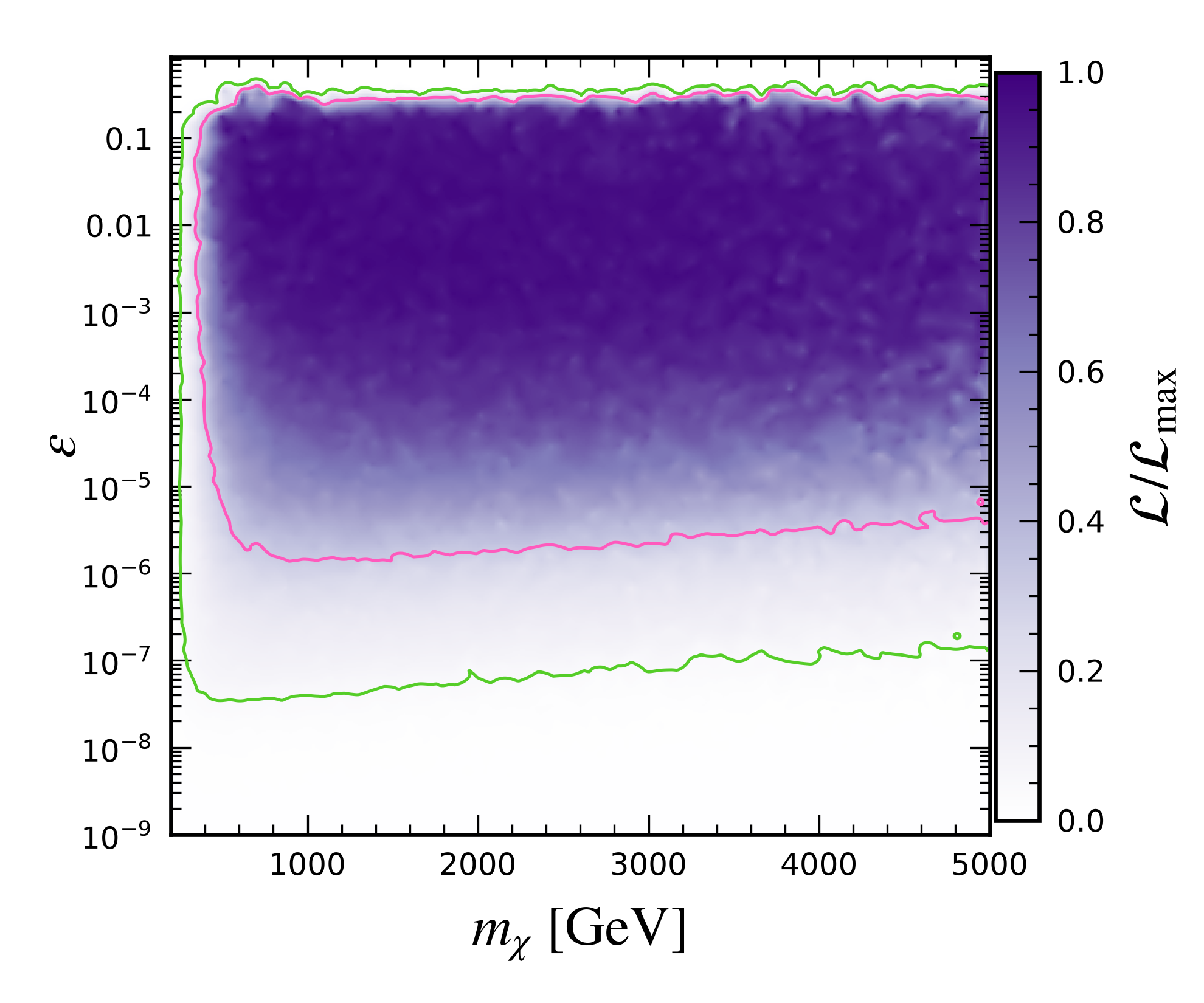}
  \includegraphics[width=0.45\linewidth]{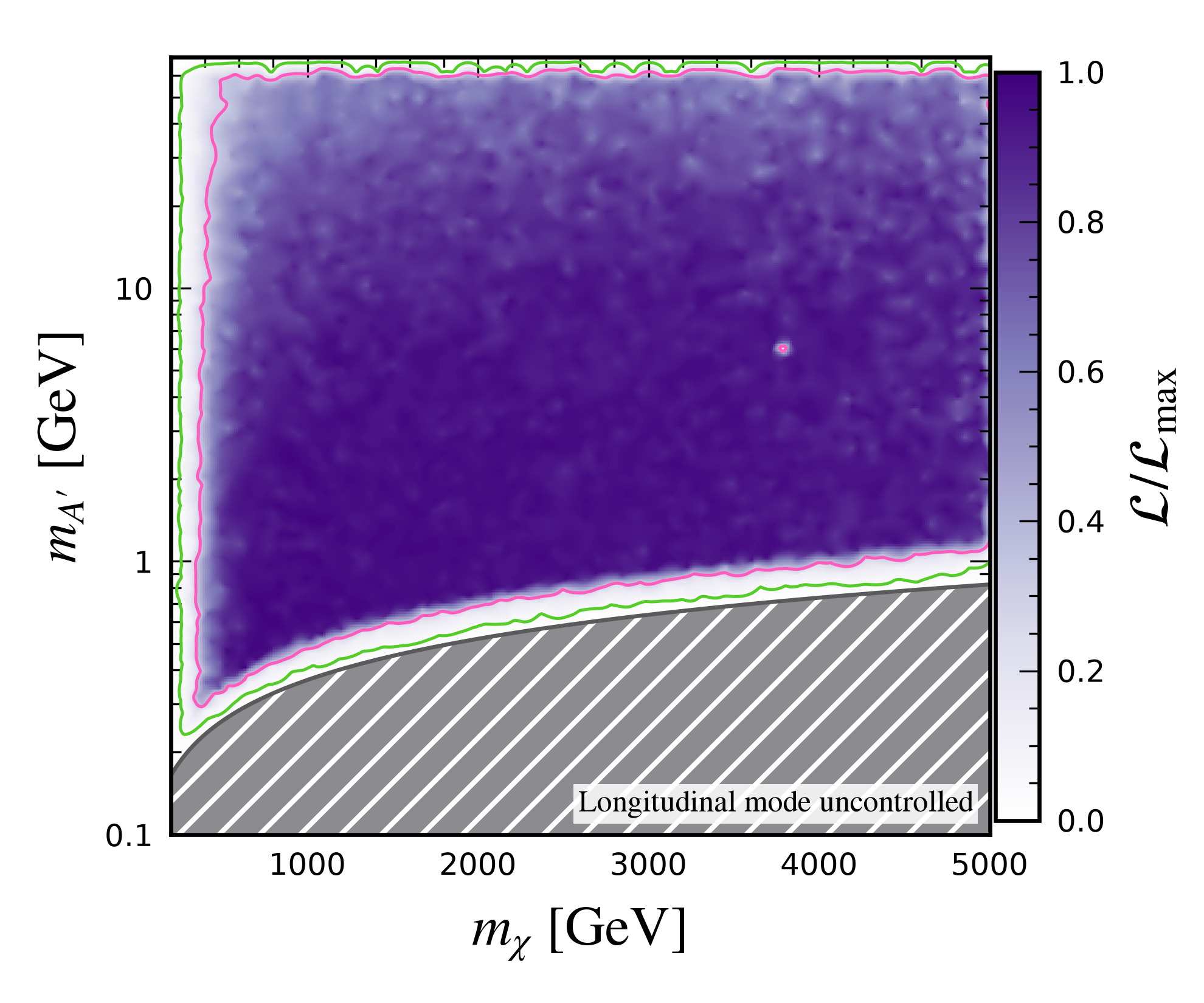}
  \caption{The profiled likelihood ratio $\mathcal{L}/\mathcal{L}_{\max}$ from the combined LZ and relic-density fit, projected onto the input-parameter planes. The pink and green contours are the $1\sigma$ and $2\sigma$ ranges. Top-left: the $(m_\chi,\delta)$ plane. Top-right: the $(m_\chi,g_D)$ plane. The line is $g_D=0.66\,(m_\chi/{\rm TeV})^{1/2}$, the locus along which secluded annihilation gives the observed relic density. Bottom-left: the $(m_\chi,\epsilon)$ plane. Bottom-right: the $(m_\chi,m_{A^\prime})$ plane. The hatched region is removed by the hard cut of Eq.~(\ref{eq:hardcut}) , where the longitudinal mode of the Proca dark photon is uncontrolled. }
  \label{fig:scan-projections}
\end{figure*}

The combined fit finds a region that reproduces both the LZ recoil candidate and the observed relic density. Figure~\ref{fig:scan-projections} shows the profiled likelihood; the pink and green contours are the $1\sigma$ and $2\sigma$ ranges. The main result is that a particular value of $m_\chi$ is not preferred: from the $200\,{\rm GeV}$ edge of the prior to $5\,{\rm TeV}$ and beyond, a pair $(m_\chi,\delta)$ can be found that fits the candidate well while maintaining $\Omega h^2\simeq0.12$. This mass independence contrasts with the LZ analysis using a DM effective operator, whose local significance depends on the assumed dark-matter mass \cite{LZ:2026highER}. Here the other parameters track $m_\chi$, so a good energy-only fit exists along the kinematic band rather than at a single mass. Table~\ref{tab:benchmarks} in Sec.~\ref{sec:rates} lists three points near the best-fit region, at $m_\chi\simeq500\,{\rm GeV}$, $1\,{\rm TeV}$ and $3\,{\rm TeV}$. Each gives $\Omega h^2\simeq0.12$ and $\mu_s\sim1$, and their $\ln\mathcal{L}_{\rm LZ}$ values differ by only $0.10$.

In the $(m_\chi,\delta)$ panel the high-likelihood band sits just below the kinematic ceiling of Eq. (15), at $\delta/\delta_{\max}\simeq0.83$–$0.92$. The Table~\ref{tab:benchmarks} points sit at $\delta/\delta_{\max}\simeq0.94$–$0.95$. Raising $\delta$ moves the recoil peak up and concentrates the signal at $E_{\rm obs}$, until the candidate energy closes. Table~\ref{tab:benchmarks} shows both an in-window peak ($E_R^{\rm peak}=239\,{\rm keV}$ for \texttt{BM-A}) and peaks above the acceptance ($334\,{\rm keV}$ and $364\,{\rm keV}$ for \texttt{BM-B} and \texttt{BM-C}). After detector response the three accepted yields and shapes at $248\,{\rm keV}$ remain comparable, so the energy-only likelihood does not select a mass.

The $(m_\chi,g_D)$ panel is fixed by the relic density. The samples concentrate on
\begin{equation}
	  g_D\simeq0.66\left(\frac{m_\chi}{1\,{\rm TeV}}\right)^{1/2},
\label{eq:relic-locus}
\end{equation}
in agreement with Eq.~(\ref{eq:sigmav}). 

The remaining two parameters are constrained only together by the rate. The $(m_\chi,m_{A^\prime})$ panel prefers a GeV-scale mediator, as required by the finite-propagator rate in Sec.~\ref{sec:xenonsr}. The lower edge of $m_{A^\prime}$ is set by Eq.~(\ref{eq:hardcut}), which removes the region where the longitudinal mode is uncontrolled; it is not a physical bound. Once $g_D$ and $m_{A^\prime}$ are set, the yield constrains the nucleon amplitude, so $\epsilon$ forms a band of $\mathcal{O}(0.01)$ in the $(m_\chi,\epsilon)$ panel, decreasing toward heavier $m_\chi$. This interval in $m_{A^\prime}$ and $\epsilon$ can be tested at colliders, as discussed in Sec.~\ref{sec:constraints}.

The corresponding zero-momentum amplitudes in Table~\ref{tab:benchmarks} are
\begin{equation}\label{eq:required-amplitude}
  \mathcal{A}_p^{\rm SI}\simeq(3)\times10^{-3}\,{\rm GeV^{-2}},
  \quad 
  \sigma_{\chi-p}^{\rm SI}\sim(5)\times10^{3}\,{\rm pb},
\end{equation} for the $\chi_1p\to\chi_2p$ transition. This large cross section is allowed because the endothermic channel is closed for most of the halo; elastic exclusion curves do not apply.

Relative to the background-only value $\ln\mathcal{L}_{\rm LZ}=-10.242$ of Eq.~(\ref{eq:logLLZ}), the Table~\ref{tab:benchmarks} points give $\ln\mathcal{L}_{\rm LZ}\simeq-5.15$ to $-5.25$. The difference $\Delta(-2\ln\mathcal{L}_{\rm LZ})$ is the likelihood-ratio improvement of this energy-only proxy over background only, and is $\simeq10.2$ at the best point. It is not a local or global significance for the LZ search.

The three benchmarks are slices through this band, not preferred masses. Whether the region is viable still depends on the late-time excited-state abundance, examined in Sec.~\ref{sec:removal}.

\section{Excited-state depletion}
\label{sec:removal}
The fit in Sec.~\ref{sec:results} takes today's relic density to reside in $\chi_1$: $\Omega h^2$ is the total two-state freeze-out abundance of Sec. 3.3, and the LZ rate is computed as endothermic $\chi_1$ scattering. Freeze-out does not give this by itself. The splitting is then negligible, so $\chi_1$ and $\chi_2$ are produced with nearly equal abundances. If a non-negligible fraction of today's relic remained in $\chi_2$, the same off-diagonal interaction would also mediate exothermic down-scattering, $\chi_2 N\longrightarrow\chi_1 N$. This channel releases the splitting energy and does not rely on the fast tail of the local speed distribution, so the rate per excited particle is very large for the LZ couplings. A recent exothermic interpretation of the LZ event occupies a different region, with substantially smaller masses and larger splittings than those preferred here \cite{deLima:2026shq}. Here the same process instead constrains the excited fraction of today's relic.

The possible depletion channels are restricted in the minimal model. Equation~(\ref{eq:la-XVint}) contains only the off-diagonal $\chi_2$–$\chi_1$–$A^\prime$ current, and the LZ kinematics require a sub-MeV splitting with $\delta<2m_e$, so the three-body decay $\chi_2\to\chi_1 e^+e^-$ is consequently closed. The remaining dominated decay channels are $\chi_2\to\chi_1\nu\bar\nu$ and $\chi_2\to\chi_1+3\gamma$, as discussed for sub-MeV inelastic dark matter in Ref. \cite{DallaValleGarcia:2025cwf}. The neutrino amplitude receives both $A^\prime$- and $Z$-exchange contributions, which cancel at vanishing momentum transfer, leaving the loop-induced three-photon channel, with characteristic lifetime
\begin{equation}\label{eq:tau-3gamma-estimate}
\tau_{3\gamma}\sim
10^{18}\,{\rm s}
\left(\frac{300\,{\rm keV}}{\delta}\right)^{13}
\left(\frac{m_{A'}}{{\rm GeV}}\right)^4
\left(\frac{10^{-3}}{\epsilon}\right)^2
\left(\frac{0.02}{\alpha_D}\right)\,.
\end{equation}
The steep $\delta^{-13}$ scaling makes $\tau_{3\gamma}$ highly sensitive to the splitting. For the Table~\ref{tab:benchmarks} benchmarks it is of order $10^{16}\,{\rm s}$, shorter than the age of the Universe, so some late decay occurs, but $3\gamma$ depletion is not an early mechanism for putting today's relic in $\chi_1$. CMB anisotropies, spectral distortions, and diffuse photons can constrain small unstable fractions in this lifetime range \cite{Balazs:2022tjl,Slatyer:2016qyl}.

Before this radiative decay acts,  number-changing processes can already reduce the relative excited abundance of excited states, in particular exothermic conversions $\chi_2\chi_2\longrightarrow\chi_1\chi_1$.   These reactions can continue after chemical freeze-out because they conserve the total dark-matter number while converting two heavier states into two lighter states. As the dark sector cools, the reverse process becomes Boltzmann suppressed; eventually the forward conversion rate also falls below the expansion rate and leaves a residual fraction $f_2^{\rm conv}$. Using the tree-level conversion rate and assuming a dark sector in thermal equilibrium with the SM, the Table~\ref{tab:benchmarks} benchmarks give $f_2^{\rm conv}$ of order $10^{-4}$ to $10^{-3}$ (similar to Ref.~\cite{Baer:2026fpy}).  Thus conversion alone does not put today's relic entirely in $\chi_1$.

We then apply the analytic $3\gamma$ lifetime to this conversion remnant. Relative to $f_2^{\rm conv}$, the present-day fractions in Table~\ref{tab:benchmarks} are exponentially suppressed, spanning from of order $10^{-6}$ at \texttt{BM-A} to $10^{-16}$ at \texttt{BM-C}. Even so, the exothermic rate per surviving $\chi_2$ remains so large that the accepted yields are of order $10^6$ events at BM-A and tens of events at \texttt{BM-B}, while \texttt{BM-C} is of order $10^{-5}$. At this level of treatment, \texttt{BM-A} and \texttt{BM-B} remain incompatible with taking today's relic to reside in $\chi_1$, whereas the faster $3\gamma$ decay of \texttt{BM-C} suppresses the present-day signal. This point-to-point variation follows from the exponential depletion together with the $\delta^{-13}$ lifetime; it cannot be represented by a single universal bound on $f_2$.

These numbers should be interpreted with an important qualification. In the preferred region, $\alpha_Dm_\chi/m_{A^\prime}$ is already large, so long-range  effects and kinetic decoupling may modify the conversion rate. We therefore use the our results only as a diagnostic of the excited-state problem and do not claim a controlled prediction of $f_2^{\rm conv}$ throughout the fitted region. A complete assessment of the minimal model would require a non-perturbative conversion calculation together with the associated cosmological limits. The $3\gamma$ width, by contrast, is the relevant perturbative decay estimate once a conversion remnant is specified.

A simple next-to-minimal possibility is a transition dipole,
\begin{equation}\label{eq:la-dipole}
	\mathcal{L}
\supset
\frac{1}{\Lambda_d}
\bar\chi_1\sigma^{\mu\nu}\chi_2 F_{\mu\nu}
+{\rm h.c.},
\end{equation}
which opens the two-body decay $\chi_2\to\chi_1\gamma$ \cite{Baryakhtar:2020rwy}. For a splitting of a few hundred keV, this decay can efficiently deplete $\chi_2$ even for a suppression scale as large as $\Lambda_d\sim10^{10}\,{\rm GeV}$,  while for LZ signals induced by this operator, one would need $\Lambda_d \sim 10^5\,{\rm GeV}$~\cite{Asadi:2026iot}. The operator couples to the photon and controls the late decay of $\chi_2$, rather than the $A^\prime$-mediated scattering $\chi_1 N\to\chi_2 N$ or the secluded annihilation $\chi\chi\to A^\prime A^\prime$ that underlie the fit in Sec.~\ref{sec:results}. It therefore need not materially alter the endothermic rate or the relic $g_D(m_\chi)$. The photon emitted in $\chi_2$ decay have the energy $E_{\gamma}\sim \delta$, of a few hundred keV. This lies below the $2.2\,{\rm MeV}$ deuterium binding energy, so the decay cannot photodissociate light nuclei and BBN does not constrain it. Energy injected after the CMB photon bath has fallen out of full thermal equilibrium can still distort the frequency spectrum. Those spectral-distortion bounds become relevant for sufficiently late decays, but still allow lifetimes of order $10^5\,{\rm s}$ \cite{Fixsen:1996nj,Acharya:2019owx}. The dipole therefore provides a concrete depletion mechanism when the minimal conversion-plus-$3\gamma$ history leaves too much $\chi_2$. 

\section{Accelerator constraints}
\label{sec:constraints}
\begin{figure}[thbp]
  \centering
  \includegraphics[width=\linewidth]{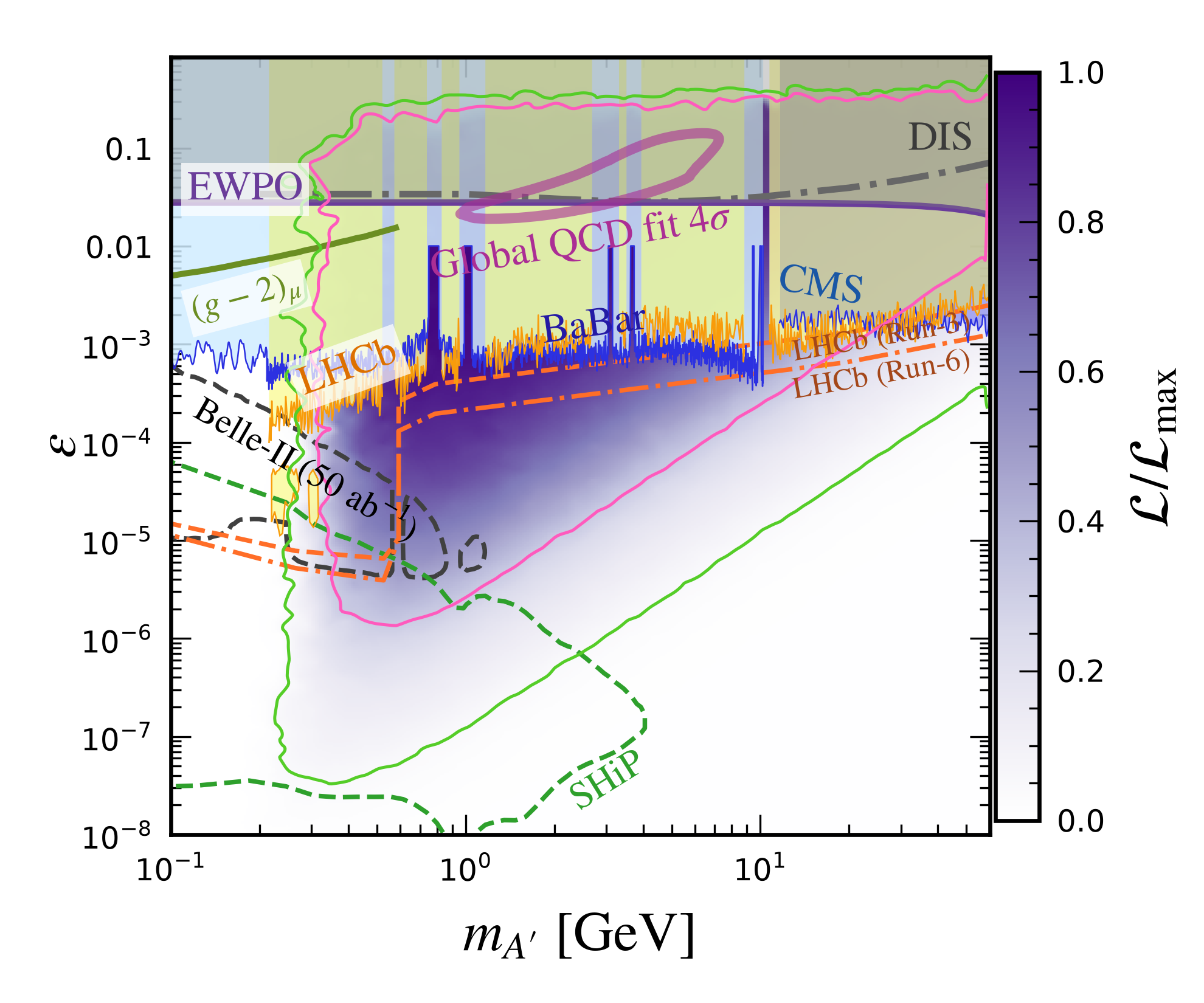}
  \caption{Profiled likelihood ratio projected onto the $(m_{A^\prime},\epsilon)$ plane, with overlaid experimental constraint contours. Accelerator constraints are from the visible BaBar search \cite{BaBar:2014zli} (solid blue), LHCb \cite{LHCb:2019vmc} (orange) and CMS dimuon search \cite{CMS:2019buh} (also blue), assuming that $A_D$ is narrow. The $95\%$-CL DIS limit obtained with refitted PDFs in ref. \cite{Thomas:2021lub} (dot-dashed grey), and the electroweak fit of ref. \cite{Curtin:2014cca} (dotted purple). The constraints from the muon $g-2$ (solid green) are taken from ref. \cite{Pospelov:2008zw}. The region above each limit is excluded. The orange contour is the preferred region of the global QCD analysis \cite{Hunt-Smith:2023sdz}; it is not an exclusion. We also show projected limits from Belle II \cite{Ferber:2022ewf}, LHCb \cite{Gori:2022vri,Craik:2022riw}, and SHiP \cite{SHiP:2020vbd}. Samples are colored as in Fig.~{\ref{fig:scan-projections}}; the scan prior extends to $\epsilon\le0.5$, while EWPO already constrains $\epsilon\sim\mathcal{O}(10^{-2})$ over much of this mass range.}
 \label{fig:constraints}
\end{figure}

The combined fit of Sec.~\ref{sec:results} prefers $\epsilon\sim\mathcal{O}(0.01)$ and a GeV-scale $m_{A^\prime}$. Existing electroweak, DIS, and collider searches already cover this $(m_{A^\prime},\epsilon)$ region; we collect the corresponding limits in Fig.~\ref{fig:constraints}.

Because of $A'$-$Z$ mixing, electroweak precision observables (EWPO) \cite{Hook:2010tw,Curtin:2014cca,Loizos:2023xbj} measured at lepton (LEP, SLC) and hadron (Tevatron, LHC) colliders \cite{ParticleDataGroup:2024cfk} constrain kinetic mixing at the level $\epsilon\sim\mathcal{O}(10^{-2})$ for GeV-scale $m_{A^\prime}$. \texttt{BM-A} in Table~\ref{tab:benchmarks} sits near or above that ballpark ($\epsilon\simeq0.06$), so these bounds must be read together with the LZ fit rather than treated as automatically satisfied. Electron–proton DIS limits \cite{Kribs:2020vyk,Thomas:2021lub,Yan:2022npz} are compatible with the EWPO bound for $m_{A^\prime}<10\,{\rm GeV}$ and weaken at larger masses. 

A recent global QCD analysis of $ep$ DIS and related data \cite{Hunt-Smith:2023sdz} has also discussed a dark-photon preference in a nearby region (e.g. $m_{A^\prime}\sim3\,{\rm GeV}$, $\epsilon\sim0.03$); we treat that as an interesting hint rather than a confirmed signal. The muon $g-2$ curves in fig. 3 are shown for comparison: they follow the older $(g-2)_\mu$ preference \cite{Pospelov:2008zw} and will be slightly stronger once the latest lattice inputs are used \cite{Aliberti:2025beg}. Note, however, that  the $g-2$ curve is well within the region already excluded by direct $A'$ searches.

The strongest direct bounds come from $e^+e^-$ \cite{BaBar:2014zli} and hadron-collider searches \cite{LHCb:2019vmc,CMS:2019buh}, as shown in Fig.~\ref{fig:constraints}. These analyses assume that the dark photon is a narrow resonance decaying only to SM final states, leading to $\epsilon\lesssim10^{-3}$ with gaps near other resonances ($\rho$, $\phi$, $J/\psi$, etc.). However, if the dark photon has a larger width from couplings to light DM (even if those channels are subdominant for the relic density), the limits can be significantly relaxed \cite{CMS:2024zqs,Felix:2025afw}.

After these limits are imposed, the $1\sigma$ region of the fit still survives within the range $0.3\,{\rm GeV}\lesssim m_{A^\prime}\lesssim 10\,{\rm GeV}$, including a triangular pocket at $\epsilon\lesssim 10^{-3}$. Proposed facilities such as FCC-ee \cite{FCC:2018evy} and CEPC \cite{CEPCStudyGroup:2018ghi} are expected to tighten electroweak constraints on the dark sector, while future measurements at Belle II \cite{Ferber:2022ewf}, LHCb \cite{Gori:2022vri,Craik:2022riw}, and the proposed SHiP experiment \cite{SHiP:2020vbd} can probe part of this remaining region. 

\section{Summary}
\label{sec:conclusions}

We have investigated whether endothermic inelastic dark matter coupled to a kinetically mixed dark photon can account for the high-energy nuclear-recoil candidate reported by LZ.
A fit of the five model parameters finds a preferred region with TeV-scale dark-matter masses, mass splittings of a few hundred keV, and a GeV-scale dark photon. The high recoil energy pushes the splitting close to its kinematic ceiling, so the scattering threshold sits in the high-velocity tail of the halo. The recoil spectrum peaks near or above the candidate energy. Secluded annihilation into dark photons sets the relic abundance and fixes the dark gauge coupling as a function of the dark matter mass. The scattering rate then selects a GeV-scale dark photon. The best-fit region predicts one accepted LZ recoil event candidate and satisfies the observed relic density. 

The same small splitting that enables the endothermic interpretation also creates an important consistency condition. The preferred splittings lie below the $e^+e^-$ threshold, and leave the excited state $\chi_2$ long-lived. Tree-level conversion and the subsequent three-photon decay process reduce the remaining excited-state population, but at lower dark-matter masses the resulting exothermic down-scattering is still too large for the minimal model to be automatically safe. A dimension-five transition dipole provides a simple next-to-minimal way to remove $\chi_2$ without changing the relic density or the endothermic LZ rate.

Finally, the GeV-scale dark photon required by the LZ interpretation is independently testable at accelerators. Existing bounds, notably electroweak precision observables and visible resonance searches, already constrain the largest kinetic mixings, but they do not close the preferred region: a surviving parameter space remains after these limits are imposed.
Future searches at Belle II and LHCb, together with the proposed SHiP experiment, can further test this remaining region.

\section*{Acknowledgements}
P.~Zhu thanks Yang Zhang for helpful discussions of the LZ results.
P.~Zhu thanks Xiaokang Du for help with the numerical calculation.
This work is partially supported by the \emph{Australian Research Council} through the ARC Centre of Excellence for Dark Matter Particle Physics (CE200100008).

\bibliographystyle{CitationStyle}
\bibliography{bibliography}

\end{document}